\documentstyle[amssymb,aps,twocolumn]{revtex}

\begin{document}
\author{F.Rullier-Albenque$^{1}$, H. Alloul$^{2}$, R.\ Tourbot$^{1}$}
\title{Disorder and transport in cuprates: weak localization and magnetic
contributions }
\address{$^{1}$SPEC, Orme des Merisiers, CEA, 91191 Gif sur Yvette, France\\
$^{2}$Laboratoire de Physique des Solides, UMR 8502, Universit\'{e}\\
Paris-Sud, 91405 Orsay, France}
\date{\today}


\twocolumn[\hsize\textwidth\columnwidth\hsize\csname@twocolumnfalse\endcsname
\maketitle

\begin{abstract}
We report resistivity measurements in underdoped
YBa$_{2}$Cu$_{3}$O$_{6.6}\,$and overdoped Tl$_{2}$Ba$_{2}$CuO$_{6+x}$
single crystals in which the concentration of defects in the CuO$_{2}$
planes is controlled by electron irradiation. Low $T$ upturns of the
resistivity are observed in both cases for large defect content. In the Tl
compound the decrease of conductivity scales as expected from weak
localization theory. On the contrary in YBa$_{2}$Cu$_{3}$O$_{6.6}$ the much
larger low T contribution to the resistivity is proportional to the defect
content and might then be associated to a Kondo like spin flip scattering
term. This would be consistent with the results on the magnetic properties
induced by spinless defects.
\end{abstract}
]

The origin of the strange metallic properties of the cuprates, among which
the transport properties, is still the key feature to be resolved to
progress in the understanding of the physics of these 2D correlated electron
systems. As the pseudogap is evidenced both in magnetic and transport
properties, some interplay between spin and charge degrees of freedom is
required. This interplay can be studied as well from the influence of
defects and impurities on the transport properties. Point defects such as Zn
impurities induce a strong depression of $T_{c}$ as expected for d-wave
superconductivity.\ This is often considered to result from pure potential
(unitary?) scattering \cite{Chien,Fuku,Tolpygo,RA}.\ However it has been
shown that such spinless defects induce a local moment in their vicinity
\cite{Mahajan,Bobroff}.\ One can therefore wonder whether these magnetic
degrees of freedom play a role in the charge scattering, and induce for
instance a (Kondo like) spin flip contribution to the resistivity.\

\smallskip In presence of Zn impurities, low $T$ upturns of the resistivity
with a $\ln T$ variation have been observed in La$_{2}$Sr$_{2}$CuO$_{4}$
(LSCO) and YBa$_{2}$Cu$_{3}$O$_{7-\delta }$ (YBCO) \cite{Ando1,Karpinska}.\
However these observations have required large Zn contents or low hole
doping in order to suppress superconductivity. Data were then limited to
large values of the 2D\ resistivity which approach or exceed $\thickapprox
5k\Omega /\square $ for which deviations from weak localization predictions
start to be observed in 2D\ systems \cite{Dolan,Bishop1}. These experimental
constraints have up to now prevented any thorough study either of the
impurity contributions to the scattering, or of the applicability of 2D weak
localization theory \cite{Abrikosov}.

\ Recent accurate NMR measurements in the case of Li impurities, which
behave similarly to Zn, have evidenced a Kondo like variation of the induced
local moment susceptibility with a Kondo temperature $T_{K}$ which increases
with hole doping \cite{Bobroff}. One might expect that magnetic
contributions to the transport properties should then differ markedly in
underdoped and overdoped systems while localization effects should always be
present.

\ In this work,\ we address both questions for the first time by comparing
underdoped and overdoped cuprates. We have used electron irradiation to
introduce in-plane point defects which behave similarly to Zn substitutions
in YBCO \cite{Legris,RA}. As measurements could be performed on a single
sample, the accuracy on the defect contribution to the resistivity was high
enough to study thoroughly its variation with defect content and
temperature, even when $T_{c}$ is still high. In the heavily overdoped Tl$%
_{2}$Ba$_{2}$CuO$_{6+x}$ (Tl2201), we find here that $T_{c}$ can be
suppressed while the low $T$ resistance is still below the strong
localization limit.\ This enables us to demonstrate that weak localization
theory applies and that defect scattering is purely elastic in this system.\
We shall show that this approach fails in underdoped YBCO$_{6.6}$. The large
magnitude of the low $T$ upturns of the resistivity and its variation with
defect content rather indicates the occurence of a Kondo like inelastic
spin-flip scattering. The difference between these two systems will be found
consistent with the observed doping dependence of $T_{K}$.

We have taken data on the underdoped YBCO$_{6.6}$ single crystal, which has
already been used in ref \cite{RA}.\ It displayed the usual pseudo gap
resistivity temperature dependence with an initial $T_{c}$ of 60K. The
highly overdoped Tl2201\ sample prepared similarly to those of ref \cite{RA}%
, with an initial $T_{c}$ of 30K, exhibited a power law dependence of $\rho
\left( T\right) \,$characteristic of the overdoped state \cite{Kubo}. The
samples have been irradiated with 2.5\ MeV electrons$^{(*)}$ in liquid
hydrogen to allow high electron flux without sample heating.\ After a short
time defect annealing ($\thickapprox $ 10 minutes) either at $T_{a}$=100K,
150K or 300K, resistivity measurements were performed between 20K and $T_{a}$%
.\ In such conditions the resistivity versus $T$ curves plotted in Fig.1
were found absolutely reversible, indicating the absence of further
evolution of the defect content. The high fluences required to reach large
enough resistivities led us to perform several independent irradiation
runs.\ This forced us to remove the probe from the irradiation facility and
to let the samples stay at room $T$ for long time (typically a few months)
after each run, which reduced subsequently the concentration of defects\cite
{RA}. The resistivity curves after annealing at 300K are quite identical to
those observed in situ for lower irradiation fluences.\ This indicates that
the measured resistivity is only sensitive to the concentration of in-plane
defects. After the last run the samples were removed from the irradiation
probe and ex-situ measurements were then performed down to lower $T$ (1.7K)
in a different set-up in which magnetoresistance data could be taken.

As can be seen in Fig.1 the high $T$ parts of the resistivity curves are
nearly parallel to each other in YBCO$_{6.6}$, i.e. Mathiessen's rule
extends to large irradiation fluences \cite{RA}, which\ implies also that
the hole doping of the CuO$_{2}$ planes is not significantly modified.\
Conversely the variation with $T$ of the resistivity of the Tl2201\ sample
increases with defect content showing that in this case the hole doping
increases slightly\cite{footnote1}.\ In both compounds low$\,T$ upturns of
the resistivity are observed and increase with defect content. These upturns
appear at higher $T$ for YBCO$_{6.6}$, and correspond to larger
contributions to the 2D in-plane resistance $R_{2D}$ than for Tl2201 \cite
{footnote2}.

To see this effect more quantitatively, we have reported in Fig.2 $\delta
R_{2D}$, the low T increase of $R_{2D}\,$obtained in these two compounds for
comparable residual resistances $R_{2D}^{*}\simeq 4.8k\Omega /\square $. In
order to evaluate $\delta R_{2D}$ the resistivity of the non irradiated
compound had to be extrapolated below $T_{c}$.\ For pure YBCO$_{6.6}$, two
different empirical forms $R_{2D}^{0}(T)=a+bT+cT^{3}$ and $%
R_{2D}^{0}(T)=a+bT^{p}\,$have been used to fit the data between 80K and
150K.\ This lower T limit has been taken to avoid the $T$ range for which
superconducting fluctuations contribute to the resistivity. Although their
extrapolations differ significantly at low $T$, the magnitude and $T$
dependence obtained for $\delta R_{2D}$ are quite identical in the limited
experimental range above $T_{c}$.\ \ For Tl 2201, superconductivity has been
suppressed for this fluence and the low $T$ extrapolation might be more
critical than for YBCO$_{6.6}$. To take into account the slight change of
hole doping we have then fitted the high $T$ part of each resistivity curve
with $R_{2D}^{0}(T)=a+bT^{p}$.\ To test the validity of this procedure, two
extreme cases have been considered. The fits were performed either between
70K\ and 150K, or between 150K and 300K.\ Although $\delta R_{2D}$ is
naturally found to depart from zero at a slightly higher $T$ in the latter
case, the overall variation and particularly the $\ln T$ dependence found at
low $T$ is not sensitive to the function chosen for $R_{2D}^{0}(T)$.\ It is
clear in Fig.2\ that, whatever the fitting procedure, $\delta R_{2D}\,$ is
larger and has a steeper $T$ variation in YBCO$_{6.6}$ than in the Tl
compound.

\smallskip We can at this stage compare these data with the expectations of
the weak 2D localization theory, which can apply as long as the elastic mean
free path $l$ verifies $k_{F}l>>1$ (where $k_{F}$ is the Fermi wave vector),
that is for $R_{2D}<<26k\Omega /\square ${\it .\ }The weak localization
contributions are expected to become sizable at temperatures for which the
elastic scattering rate $1/\tau $ becomes much larger than the inelastic
scattering rate$\,1/\tau _{i}$.$\;$Assuming that the in-plane defects only
contribute to the elastic scattering, the ratio $\tau _{i}/\tau \,$ can be
estimated from the data, if the inelastic rate is associated to the $T$
dependent contribution to $R_{2D}$ of the starting pure material.\ For
instance, for the residual resistivity $R_{2D}^{*}\simeq 4.8k\Omega /\square
$ of Fig.2, $\tau _{i}/\tau \ $ $\thickapprox 10-12$ and $\thickapprox 7$
respectively for Tl2201 and YBCO$_{6.6}$ at 100K ( at 50K\ the deduced value
would be $\tau _{i}/\tau $ $\thickapprox $40 for the two systems).\
Therefore, within the above assumptions, weak localization conditions are
fulfilled even at such high $T$.\ Let us recall that the theoretical
correction to the 2D conductance has a universal behaviour at low $T$ given
by

\begin{equation}
\Delta \sigma \,\,=\frac{\alpha e^{2}}{2\pi ^{2}\hbar }p\ln (T/T_{0})
\label{eq.1}
\end{equation}
where $p\;$is the exponent of the $T$ dependence of the inelastic scattering
rate$\,\,1/\tau _{i}\varpropto T^{p}\,\,,\;$and\ $\alpha =1\,$\ in the
absence of electronic interactions, in good agreement with experimental
observations \cite{Lee,Bishop2,Bergmann,Abraham}{\em .}

\ We have plotted in Fig.3a the value of $\Delta \sigma
=(R_{2D})^{-1}-(R_{2D}^{0})^{-1}$ for the Tl compound.\ For the in-situ
measurements the data are limited to 20K, while they could be pursued down
to 3K for the final state of the sample.\ In this case a good fit with a $%
\ln T$ dependence extends over a decade.\ One can notice in Fig.3a that the
curves for the other defect contents\ are parallel to each other at 20K,
with a slope identical to that found for the lower $T$ run, within
experimental accuracy.\ Finally at a given $T$ the magnitude of $\Delta
\sigma $ increases with defect concentration, as the upturn in the
resistivity curve occurs at higher T as can be seen in Fig.1a. These
qualitative observations are more striking in Fig.3b where we have performed
the scaling of these curves in{\em \ }$\ln (T/T_{0})$.\ \ Here $T_{0}$\ is
found to increase progressively from 70K\ to 120K with defect content. The
actual slope obtained experimentally below 30K yields the numerical value $%
\alpha p=1.3,$ when compared to (\ref{eq.1}). The value of $p$ can be
independently estimated from the $\,T$ dependence of $R_{2D}^{0}(T)$ and
varies from 1.7 to 2.3. This would correspond to $\alpha \simeq 0.7.\,$These
results provide a strong experimental evidence that 2D\ weak localization
theory explains the Tl2201 data.

Transverse magnetoresistance measurements performed up to 8T after the last
irradiation provide an independent evidence which confirms this analysis. It
indeed exhibits{\it \ a negative contribution }which decreases with
increasing T as expected for weak localization.{\em \ }The field variation
was accurate enough at low $T$ to allow an analysis along the lines of ref.
\cite{Bishop2}, which allowed to deduce $\alpha \simeq 0.5\,$ and\thinspace $%
\,\tau _{i}/\tau \simeq 60\;$at $4.2\;K.$ While the value of $\alpha $ is in
good agreement with the $T$ dependence of $R_{2D}$, the value of $\tau
_{i}/\tau \ $is smaller than that which would be obtained from the
extrapolation of the $T$ dependent part of the resistivity of the pure
sample.\ Such a trend is quite common, as $\tau _{i}$ is usually found to
flatten at low $T$, at values smaller, even in order of magnitude, than
expected from the pure systems estimates \cite{Lee,Bergmann,Markiewicz}.
Further experiments are needed to determine whether interaction effects
contribute to $\Delta \sigma $ \cite{Larkin}.

Let us consider now the data for the underdoped compound YBCO$_{6.6}$, for
which we have seen that $\tau _{i}$/$\tau $ is weaker or similar, whereas $%
\delta R_{2D}$ is larger than for the overdoped compound.\ In this case
superconductivity prohibits to investigate the low $T$ range, and weak
localization contributions are not expected to be fully developed.\ However,
as can be anticipated from Fig.2, a plot similar to that of Fig.3 gives much
larger values of $\Delta \sigma $ than those found for Tl2201.\ The slope
found at the lowest $T$ for which $T/T_{0}$\ is only 0.5 would extrapolate
at low $T$ to a value $\alpha p>5,$ which is absolutely not sound
physically. We could obtain a similar conclusion for electron irradiated YBCO%
$_{7}$ for which $p=1$ for the well known $T$ linear dependence of $\rho (T)$%
\cite{YBCO7}. Therefore all these data cannot be explained by weak 2D\
localization theory, as was also concluded for Zn substitution in YBCO$%
_{6.7} $ \cite{Ando1}.\

Alternatively, we can wonder whether spin flip scattering effects can
account for these large low T contributions.\ While for weak localization $%
\delta R_{2D}$ scales with $R_{2D}^{2}$ from (\ref{eq.1}), a Kondo like
contribution to $R_{2D}$ is expected to scale linearly with the defect
content, that is with $R_{d},$ the increase of the residual resistivity.
Indeed one can see in Fig.4 that $\delta R_{2D}/R_{d}\,\,$follow the same $T$
dependence for a large range of $R_{d}$ values, if one omits to consider the
dowturn associated with the proximity of the superconducting transition.
Upward deviations with respect to this common behaviour begin to occur at
low $T$ for$\,R_{d}\gtrsim 3.8k\Omega ,$ that is $R_{2D}\gtrsim 5k\Omega $.
This is even more prominent for the largest defect content in Fig. 1b which
varies faster than $\ln T$, and might therefore correspond to the onset of
strong localization.

It is interesting at this stage to compare these results with that obtained
for underdoped Zn substituted samples \cite{Ando1}.\ In that case refined
analyses, as done here, are impossible in the high $T$ range as the
substituted sample has to be compared to a pure sample which is physically
distinct.\ Systematic errors due to differences in the sample geometries and
hole contents can then be hardly avoided.\ However superconductivity could
be nearly suppressed with a 18 T field in YBCO$_{6.7}$\ with 2.7\%Zn \cite
{Ando1} which allowed to evidence a $\ln T$ contribution to the resistivity
(not to the conductivity) below 30K. This would be compatible with a Kondo
contribution with a low $T_{K}$ value.\ A rough estimate for $\delta
R_{2D}/R_{d}$ yields a value for this $\ln T$ contribution about a factor
two smaller than that obtained here. This order of magnitude agreement is
quite satisfactory\cite{footnote3}, as we are comparing here the influences
of different microscopic defects.

In conclusion, we have evidenced that the low $T$ resistivity behaves quite
differently in overdoped and underdoped systems.\ In the former case we have
shown the relevance of weak localization theory, as $\Delta \sigma $ has a $%
\ln T$ variation independent of defect content and in rough quantitative
agreement with the exponent of $R_{2D}(T)$.\ A negative transverse
magnetoresistance has been evidenced at low $T$, but further experiments are
needed to determine the eventual contribution of interactions to the $\ln T$
term and the low $T$ variation of $\tau _{i}.$

We have shown that the large $\delta R_{2D}$ measured in the underdoped case
can be interpreted in terms of a Kondo scattering on the local moment
induced by the ''spinless'' defects. This implies {\it per se} that the
scattering by such defects should saturate at low $T$ \cite{footnote4} and
should only become unitary below $T_{K}$.\ Therefore, pure potential
scattering by spinless impurities is only expected for $T<<T_{K}$. As large $%
T_{K}$ values, exceeding 200K, have been evidenced by NMR\ only for the
overdoped case \cite{Bobroff}, pure elastic scattering is expected at low $T$
for these systems. This is then quite coherent with the occurence of weak
localization contributions in the Tl2201 compound, if the NMR observations
for Li impurities do apply as well for irradiation defects.

It is worth pointing here that -$\ln T$\ contributions to $R_{2D}$ were also
observed for the ''pure'' underdoped LSCO \cite{Ando3} and Bi(SrLa)CO 2201
\cite{Ando2}.\ They have been associated with a ''peculiar'' charge
localization leading to a metal insulator transition (MIT) with decreasing
hole content. It is usually assumed that some disorder is responsible for
the anomalously low value of the optimum $T_{c}$ in these compounds and that
this disorder might be at the origin of the -$\ln T$ term. Our data lead us
to suggest that this unknown disorder might result primarily in a scattering
on magnetic perturbations induced in the planes, as for spinless impurities.
In both cases, one can wonder whether the scattering saturates at low $T$ or
whether strong localization proceeds before this saturation can be detected.
In other words, the MIT might result from the conjunction of uncontrolled
disorder and of the enhanced sensitivity of the underdoped regime to this
disorder. This would explain why the MIT is strongly dependent on the
cuprate family.

We thank P.\ Lejay and A.\ Tyler for providing the single crystals and the
technical staff of the LSI for their support during irradiation experiments.

\begin{figure}[tbp]
\caption{The measured in plane resistivity $\rho $, scaled by the lattice
constant $c$ and the number $n$ of CuO$_{2}$ planes to give the 2D
resistance $R_{2D}=\rho n/c$, is plotted versus temperature for different
irradiation conditions. The arrows indicate the sequence of measurements:
increasing irradiation for straight arrows and long annealing at room $T$
for wiggled arrows. a)Tl2201 : the data down to 1.7K has been taken in a
different set up. b)YBCO$_{6.6}$ : the extrapolated low $T$ variation of R$%
_{2D}^{0}(T)$ is shown as a full curve}
\label{fig1}
\end{figure}

\begin{figure}[tbp]
\caption{The low T increase $\delta R_{2D}$ of the resistivity is plotted
for $R_{2D}$ of about $4.8\ k\Omega /\square $ for the two compounds.\ The
different fitting procedures used to estimate $\delta R_{2D}$ are described
in the text: for YBCO$_{6.6},\,$fits with $a+bT^{p}$ or $a+bT+cT^{3}$(a)$.$
For Tl2201, fit with $a+bT^{p}\,$between 70K and 150K (b) or between 150K
and 300K (c). }
\end{figure}

\begin{figure}[tbp]
\caption{a) Plot of $\Delta \sigma $ versus ln$T$ for different defect
concentrations in Tl-2201. b) These curves are scaled by adjusting $T_{0}$
in\ equ.\ (\ref{eq.1}).}
\end{figure}

\begin{figure}[tbp]
\caption{The low $T$ increase of $R_{2D}$ is normalized to the defect
contribution to the residual resistivity $R_{d}$ in YBCO$_{6.6}$. Except for
the onset of superconductivity, the data are found to collapse onto a unique
curve as long as $R_{d}\lesssim $\ $4k\Omega /\square $.}
\end{figure}

\smallskip


\begin{references}
\bibitem{footnote1}  The fluences used for the determination of the relation
between $T_{c}$ and $\Delta \rho $ in ref \cite{RA}, are much smaller than
those required in the present experiment.\
\bibitem{footnote3}  Let us point out that the large applied field might
even reduce the Kondo contribution in those experiments.
\end{references}
\end{document}